# IO-JUPITER SYSTEM: A UNIQUE CASE OF MOON-PLANET INTERACTION

**Anil Bhardwaj**[1] **and Marykutty Michael** [2]

[1]Space Physics Laboratory, Vikram Sarabhai Space Centre, Trivandrum 695022, India, spl_vssc@vssc.org
[2]Engineering Physics, University of Virginia, Charlottesville, VA 22904-4238, USA, mm2eq@virginia.edu

## ABSTRACT

Io and Jupiter constitute a moon-planet system that is unique in our solar system. Io is the most volcanically active planetary body, while Jupiter is the first among the planets in terms of size, mass, magnetic field strength, spin rate, and volume of the magnetosphere. That Io is electrodynamically linked to Jupiter is known for nearly four decades from the radio emissions. Io influences Jupiter by supplying heavy ions to its magnetosphere, which dominates its energetic and dynamics. Jupiter influences Io by tidally heating its interior, which in turn drives the volcanic activity on Io. The role of Io and Jupiter in their mutual interaction and the nature of their coupling were first elaborated in greater detail by the two Voyagers flybys in 1979. Subsequent exploration of this system by ground-based and Earth-satellite-borne observatories and by the Galileo orbiter mission has improved our understanding of the highly complex electrodynamical interaction between Io and Jupiter many fold. A distinct feature of this interaction has been discovered in Jupiter's atmosphere as a auroral-like bright emission spot along with a comet-like tail in infrared, ultraviolet (UV), and visible wavelengths at the foot of Io flux tube (IFT). The HST and Galileo and Cassini imagining experiments have observed emissions from Io's atmosphere at UV and visible wavelengths, which could be produced by energetic electrons in the IFT. In this paper an overview on these aspects of the Io-Jupiter system is presented, which by virtue of the nature of its electrodynamical coupling, has implications for the extra-solar planetary systems and binary stars.

## INTRODUCTION

Io-Jupiter system bears a unique distinction in our solar system. While Io is the most volcanically active planetary body, the Jupiter is the largest planet, has strongest magnetic field, fastest spin, biggest and most powerful magnetosphere, and the densest of the planetary atmospheres. The study of this distinct moon-planet system is important because it is very different from the Lunar-Earth relationship, it helps advance our understanding of basic plasma-neutral-surface interactions, and it has implications to understand similar processes occurring at other places in our solar system and in extra-solar planetary systems.

Jupiter is 5.2 times farther from the Sun compared to Earth, and has an equatorial radius of 71492 km, which is 11 times (1400 times in volume) that of the Earth. Jovian atmosphere is mostly made of the simple molecules of hydrogen and helium with sulfur, oxygen and nitrogen in small amount. Jupiter's magnetic moment is about 4.3 Gauss-$R_J^3$, which is 20,000 times greater than that of Earth, with magnetic field direction opposite to that on Earth and inclination of 9.6°, which is close to 11° tilt on the Earth. The general form of Jupiter's magnetosphere resembles that of Earth with dimensions about 1200 times greater, as the solar wind pressure at 5.2 AU is only 4% of its value at 1 AU. Note that even though Jupiter is only slightly bigger than Saturn, its magnetic field is 4 times as big in each dimension. While the magnetic field of Earth is generated by the iron core, the Jovian magnetosphere is generated by the motion of magnetic material inside the liquid metallic shell. At about 1000 km below the cloud top the hydrogen atmosphere becomes thicker and finally changes phase to become liquid hydrogen. Under the liquid hydrogen layer a layer of metallic Hydrogen exists which causes the Jovian magnetic field. Unlike a dipole at Earth's core, a quadrupole and octupole also contribute to produce the Jovian magnetic field, which explains the shape of its magnetosphere. Jupiter, with a 10-hour day, is the fastest rotator among the planets. The power for populating and maintaining the magnetosphere of Jupiter comes principally from the rotational energy of the planet and the orbital energy of Io, whereas the power source for Earth's magnetosphere is principally the solar wind.

Jupiter and its satellites constitute a miniature solar system. Galilean satellites are the important among them, which were discovered by Galileo Galilei in 1610. Io, the inner most among them (while Europa, Ganymede and Callisto are the others in order of their distance from Jupiter), has a radius of 1820 km, which is only 2% larger than that of the Moon. Io orbits Jupiter at a distance of 5.9 $R_J$ away from Jupiter while Moon is at a distance of 60 $R_E$ from Earth. Io orbits within Jupiter's intense magnetic field while the radius of the Earth's distant magnetic tail occurs at about 30 $R_E$. That means while Io constantly couples with Jupiter's magnetosphere, Moon does not. Io is a volcanically active body due to the tidal heating produced by Jupiter and its orbital resonance with the other Galilean satellites, while the volcanoes on Moon have occurred between 3-4 billion years ago. The volcanoes on Io emit $SO_2$ rich material that is one of the major sources of $SO_2$ atmosphere on Io. Moon has a very thin atmosphere with He, Ar, Na, K, Ne etc as the constituents. A detailed review on Io is presented in [1].

The nature of an atmosphere on Io is of great scientific interest ever since the discovery of the Ionian ionosphere by the Pioneer spacecraft in 1973. $SO_2$ has been discovered in Io's atmosphere in infrared by Voyager, in ultraviolet by HST and Galileo, and in millimeter wave by ground-based observations [2,3,4,62,63]. The presence of SO has been observed by [5] in the millimeter wave rotation lines and $S_2$ in ultraviolet by [6]. Recently, the presence of $SO_2$, SO and $H_2S$ in the exosphere of Io has been inferred from Galileo magnetometer data [7]. Na, K, and Cl in trace amounts are also present in Io's atmosphere [1,8,9].

The sources of $SO_2$ in the atmosphere of Io are volcanic activity, sunlit sublimation of surface $SO_2$ frost, and sputtering. The major sources are the former two, but the relative importance between them is poorly understood. Sublimation produces a uniform atmosphere, whereas volcanic activity is responsible for the spatial variation or patchiness of the atmosphere of Io. A number of models have been developed to study the sublimation driven, volcanic, and sputtered atmosphere of Io [cf. Ref. 2 for review]. The important photochemical models that describe the vertical structure of the atmosphere of Io are those given by [10,11,12].

As mentioned earlier, Io orbits inside the magnetosphere of Jupiter, and therefore the atmosphere of Io is in constant interaction with the magnetospheric plasma. In the present paper the interaction of Jovian magnetospheric plasma with the atmospheres of Io and Jupiter is discussed in context of Io's atmopsheric emissions and emisions from the footprint of Io flux tube in Jupiter's atmosphere. The application of the Io-Jupiter system study to the astrophysical and cosmic objects, like binary stars and extra-solar systems, is also discussed.

## IO-JOVIAN MAGNETOSPHERE-IO PLASMA TORUS CONNECTION

Sporadic radio waves of 22.2 MHz (decametric wavelengths) and constant radiation of frequency from 300 to 3000 MHz (decimetric wavelengths) are observed from Jupiter in the late 1950s. The electrodynamic relation between Jupiter and Io was discovered when [13] noticed that these emissions are correlated with the Io's orbital position. Steady state models were developed initially by [14] and [15]. The orbital velocity of Io is less than the rotational velocity of the Jovian magnetospheric plasma. Therefore the Jovian field lines and the associated plasma stream past Io at a relative speed of 57 km s$^{-1}$, which sets up a potential difference of ~400 kV across Io resulting in a current of ~$10^6$ A that closes through the Jovian and Ionian ionospheres. The existence of such a current system was verified with the observations made by Voyager 1 [16]. The interaction of magnetospheric particles with the Io's atmosphere and surface causes sputtering and the sputtered ions are picked up by the magnetic field lines and get trapped. Hence, Io's orbital path is populated with $S^+$ and $O^+$ ions (photochemical products of $SO_2$), which is known as the Io plasma torus. The Io plasma torus is a manifestation of the relation between Io and Jupiter's magnetosphere. Na, K and Cl are also found in the torus [1,8,9]. The coupling between the plasma torus and the Io's atmosphere is studied by many groups, the recent one is by [17]. The bombardment of Jovian magnetospheric particles onto the Io's surface can cause generation of soft x-rays, as recently discovered by Chandra X-ray Observatory [18].

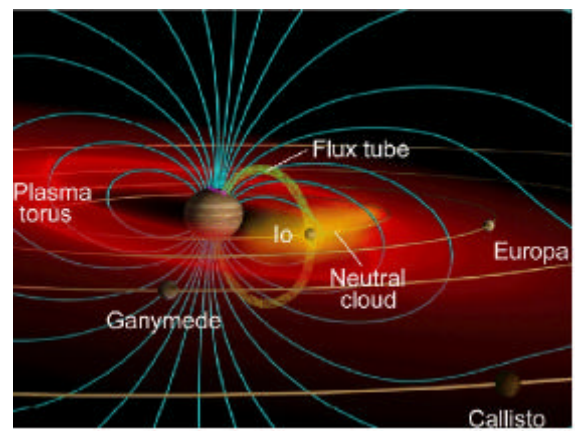

Fig. 1. A schematic showing the Jupiter, Io, Jovian magnetic field lines, Io plasma tours, and its other three Galilean satellites (not drawn to scale).

Recent Galileo observations have improved our knowledge about the Io-Jupiter relation considerably. The particles and field instruments detected strongly perturbed fields, beams of energetic electrons and ions, and a dense, cold decelerated plasma flow in Io's wake [19,20,21,22]. After having several observations by the Galileo magnetometer, [23] argued that the uncertainty regarding the Io's magnetic moment cannot be eliminated. From the asymmetries observed by the Galileo plasma wave instrument during the various flybys [24] suggested that the Io's ionospheric plasma density is being strongly influenced by the magnetospheric plasma flow around Io, similar to the radio occultaion experiment observations reported by [25]. Detection of emissions at the Io footprint in Jupiter's auroral atmosphere in the infrared, ultraviolet, and visible walengths [cf. review by Ref. 26] revealed that the particles associated with Io reach Jupiter. The interactions of magnetospheric plasma with Io's and Jupiter's atmospheres produce emissions, which are discussed in the next sections.

## MULTI-WAVELENGTH FOOTPRINTS OF IO FLUX TUBE IN JUPITER'S ATMOSPHERE

As stated above, the Io is involved in a complicated electrodynamical coupling with its plasma torus and Jupiter's magnetosphere. The electrodynamical interaction between Io and the Jovian magnetosphere results in an electric circuit that runs from Io along Jupiter's magnetic field lines and closes through the Jovian ionosphere (near 65° north and south latitude) at each foot of the Io Flux Tube (IFT). Where the particles carrying this current impact the atmosphere of Jupiter, an auroral-like spot of emission results.

The first direct evidence of the IFT footprint was obtained in a near-infrared image of Jupiter's $H_3^+$ emissions at 3.4 µm (cf. Fig. 2) in 1992 [27]. Subsequently, the IFT signature has been observed at far ultraviolet, by HST (cf. Fig. 3), and visible (by Galileo SSI) wavelengths [28,29, 30,31,32,33; cf. Ref. 34, 26 for reviews). Part of the Jovian decameter radio emissions, and especially the "S-bursts", are believed to be directly related to the IFT. The emitted power in IFT footprint is ~$0.5\times10^{11}$ W in IR ($H_3^+$ emissions), $<$~$10^{11}$ W in FUV, ~$5\times10^8$ W in visible, and ~$10^{8-10}$ W in radio wavelength ranges [34]. The radio footprints are low-frequency (~2-40 MHz) radio emissions with specific time-frequency characteristics, both L and S bursts, that originate a few thousand kilometers above the optical spots [35,38].

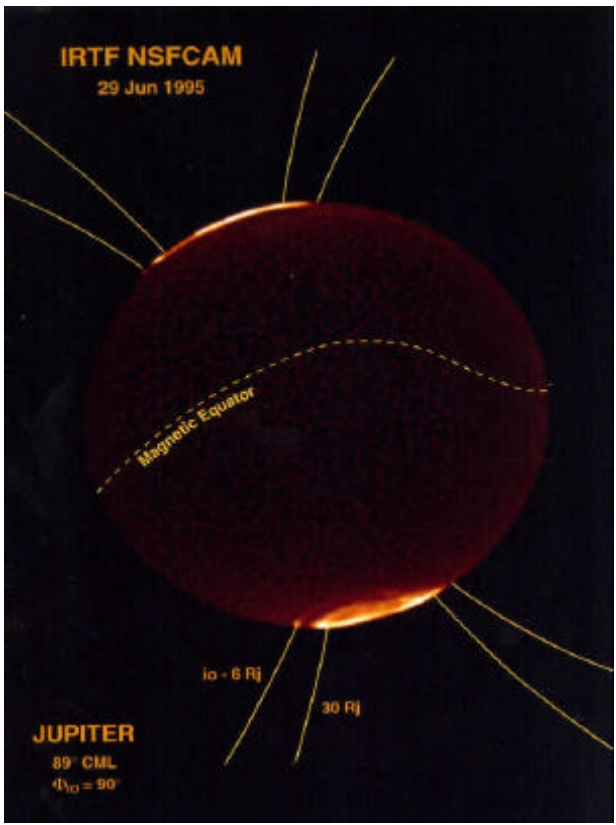

Fig. 2. Image of Jupiter at 3.4 µm obtained at NASA-IRTF. The distinct faint emission feature at the foot of L=6 $R_J$, seen in both N and S poles, is the IFT footprint. The footprint is more clearly visible in the southern hemisphere along with a trail of weaker emission extending tens of degrees in longitude downstream of the Io footprint.

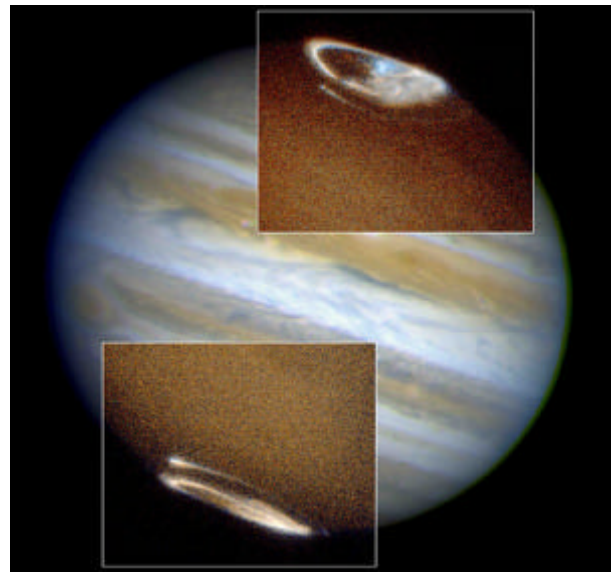

Fig. 3. HST-STIS false-colour image of Jupiter in FUV. The spot emission features, both in the northern and southern polar regions, equatorward of the main auroral oval are the IFT footprints. A weaker tail-like feature extending several tens of degree along the L=5.9 shell associated with both N and S spots are also clearly seen. (from STScI web site http://oposite.stsci.edu/pubinfo/1998.html).

The total power of multi-wavelength radiation emitted from the IFT footprint is ~$10^{11}$ W, which is <1% of the total auroral emission output of Jupiter [26]. The two main features of the "optical" IFT footprints are:

1. A spot of emission having an elliptical or roughly circular shape. The dimension of the IFT footprint is typically larger than the size of the Io projected on to the planet (since the magnetic flux tubes become narrower with decreasing distance to Jupiter, the 3640 km diameter of Io projects an ellipse of ~120×200 km at Jupiter's atmosphere). This indicates that the region of Io-Jupiter interaction is somewhat larger than the size of Io.

2. A faint comet tail-like feature associated with the main footprint has also been observed at UV, visible, and IR wavelengths (cf. Figs. 2,3). This trail of emission follows the Io's magnetic latitude in the downstream direction along the L=5.9 footprint line of magnetic latitude, and its brightness decreases slowly with increasing distance from the main footprint. The trail is observed to extend to 100° in longitude. This implies that Io's magnetic footprint extends well beyond the instantaneous magnetic mapping of Io and persists for a few hours after Jupiter's magnetic field has swept past Io.

These faint auroral trails are seen when Jupiter's dipole is tilted towards Io. In this orientation, Io is at its highest Jovicentric latitude and is temporarily outside of its plasma torus. There is no perfect explanation for the occurrence of the faint trails associated with the main footprint. At present they seem to be more consistent with the multiple reflecting Alfvèn wave model [36,37,38], but instead of discrete spots continuous emissions are seen. Within the observational sensitivity of the HST-STIS, no emission extending in the upstream direction from Io is detected in FUV [30]. Recent study [39] indicates that the FUV IFT footprint emissions are excited at higher altitudes compared to the main polar aurora and also suggests that drop in the brightness of emissions along the Io trail is due to decrease in the electron number flux rather than softening of the electron mean energy. The average energy of the electron producing the main FUV footprint is estimated to be ~50 keV.

The IFT footprints are less visible in the northern hemisphere over longitudes were the surface magnetic field is stronger (compared to the conjugate point in the southern hemisphere). This is true because the precipitation is easier in lower magnetic field regions (because of lower mirror point altitudes) compared to higher B-field region. Conversely, radio emissions are more intense over longitudes where the surface magnetic field is stronger, because radio emission generation requires substantial (unstable) reflected electron population. The emission from the IFT footprints provide a very useful "fiducial mark" for evaluating models of Jovian magnetic field, particularly close to the planet, since although the longitude of the emission may vary according to the dynamics of the IFT, the latitude of the "optical" spot maps out the L-shell of Io at 5.9 $R_J$. This wonderful constraint has helped the development of a new model (VIP4) of Jupiter's magnetic field [40] which is not yet perfect, but much improved for interpreting IFT radio emission morphology [38].

## EMISSIONS FROM IO'S ATMOSPHERE

Emissions of atomic oxygen and sulfur from Io in the ultraviolet region were observed initially by [41] using the IUE. The FOS and GHRS instruments on HST have also observed ultraviolet emissions from Io [42]. The HST-STIS has sufficient resolution to study the spatial structure of the UV emissions in detail [43,44,45]. Enhanced UV OI and SI emissions above Io's limb near its equator, called equatorial spots, are observed at sub- and anti-Jovian positions that move with the changing orientation of Jupiter's magnetic field and appear uncorrelated with the locations of the volcanic vents: suggesting the relation between Io's atmosphere and the Jovian magnetosphere.

The visible emissions of Io are observed by Voyager, ground-based telescopes, Galileo, HST, and Cassini. Visible emissions were observed for the first time by Voyager 1 spacecraft [46]. From 1990 onwards the stellar echelle spectrograph at Kitt peak has been observing OI 630 nm emissions from Io [47,48,49]. The OI 630 nm emissions are also observed by HST WFPC2 and STIS [cf. Ref. 44,49]. The visible emissions of Io from Galileo observations are reported by [50,51,52]. These visible emissions consist of blue, red, and green band emissions. Imaging sub-system on Cassini has observed visible emissions from Io in late 2000 and early 2001 on its way to the Saturn [53]. The Io's visible emissions are also found to be correlated with the orientation of the Jovian magnetic field, thereby reinforcing the relation between Io and the Jupiter's magnetic field.

The probable sources of UV and visible emissions from Io are the excitation of atmospheric neutrals by photoelectrons, torus plasma electrons, and Jovian magnetospheric electrons. Ref [42] have ruled out resonant scattering and recombination as the possible sources since the contributions from them are too low. Ref. [54] developed a Monte Carlo model to calculate the photoelectron-excited intensities of neutral and ionized sulfur and oxygen UV emissions, and SO and $SO_2$ band emissions in the wavelength region 240-430 nm, from the atmosphere of Io and found that the calculated brightness are too low to account for the HST-observed brightness.

Ref. [55] described the interaction of torus plasma electron flux with the atmosphere of Io using a three-dimensional two-fluid plasma model. This work assumes a $SO_2$ atmosphere with a scale height of 100 km and column density $3 \times 10^{16}$ cm$^{-2}$ and an atomic oxygen mixing ratio of 0.1. Though this model could explain the radiation patterns observed by HST-STIS, the observed brightness could be explained only if the mixing ratio of oxygen relative to $SO_2$ was increased to 0.2 in the upper atmosphere of Io.

Intense beams of energetic (>15 keV) electrons propagating parallel and anti-parallel to the magnetic field have been observed by the Galileo Energetic Particle Detector (EPD) [20,21,22]. Ref. [56] have reported the detection of large fluxes of bi-directional magnetically field aligned electrons at lower energies (0.1-5 keV) by the Galileo Plasma Science (PLS) instrument. The energy flux of low energy electrons is ~2 erg cm$^{-2}$ s$^{-1}$, which is about 2 orders of magnitude larger than the energy flux of high energy electrons observed by the EPD. The penetration of these electrons into the atmosphere of Io and subsequently the production of emissions are investigated by [57] using a Monte Carlo Model [58]. The brightness is calculated for two model atmospheres of Io developed by [10,11,12]. The model results showed that the electron spectrum observed by the Galileo-PLS in the energy region 0.1-5 keV are capable of producing the HST-STIS observed emissions if a fraction of these electrons precipitate into the atmosphere of Io. Table 1 presents the intensities of emissions calculated using the electron flux observed by PLS instrument and HST-observed intensities of certain important emissions.

Ref [59] used the Monte Carlo model [58] to calculate the brightness of emissions in the visible region observed by the Galileo-SSI. The visible emissions observed by Galileo-SSI consist of blue (380-440 nm), red (615-710 nm), and green (510-605 nm) bands. The probable sources of these emissions are $SO_2$ band in the blue region, OI (630 and 634 nm) in the red region, and OI (557.7 nm) and Na (589 and 589.6 nm) in the green region. The atmospheric models developed by [10,11] are used in the study and the preliminary results from this study are presented in Table 1. These intensities are calculated using the electron energy spectrum observed by the PLS instrument onboard Galileo. It is seen from the table that the red and blue emissions could be produced by the interaction of the PLS-observed field-aligned electrons with the atmosphere of Io, but this process could produce only 25% of the observed green emissions. Further results will be published soon with an upgraded model and improved input data.

Table 1. Model calculated and HST-STIS-observed FUV and Galileo-SSI-observed visible emissions on Io

| Features | Calculated (kR) | | Observed (kR) |
|---|---|---|---|
| | WJ96 | SS96 | |
| OI (130.4) | 5.9 | 20.5 | [a]1.0 |
| OI (135.6) | 1.84 | 7.3 | [a]0.9 |
| SI (147.9) | 1.44 | 1.0 | [a]1.2 |
| Blue | 11.8 | 6.2 | [b]8.2 |
| Red | 2.1 | 4.0 | [b]6.8 |
| Green | 1.7 | 2.0 | [b]8.0 |

WJ96 = Wong and Johnson (1996) [Ref. 10]; SS96 = Summers and Strobel (1996) [Ref. 11] model atmospheres.
The wavelengths in parentheses of the features are in nm.
[a]Observed by HST-STIS. [b]Observed by Galileo-SSI.

## IMPLICATIONS FOR BINARY STARS AND EXTRA-SOLAR SYSTEM

An analogy of the Io-Jupiter system is now being applied to the astrophysical and cosmic objects. A conducting body traversing a magnetic field produces an induced electric field. When the circuit is closed, a current will set up, resulting in resistive dissipation. The Jupiter-Io system therefore operates as a unipolar inductor. Another potential cosmic unipolar inductor could be a planet orbiting around a magnetic white dwarf [60]. These systems have a similar configuration, with the differences being their orbital period and separation, the masses and radii of the two components, and the magnetic moment of the magnetic body.

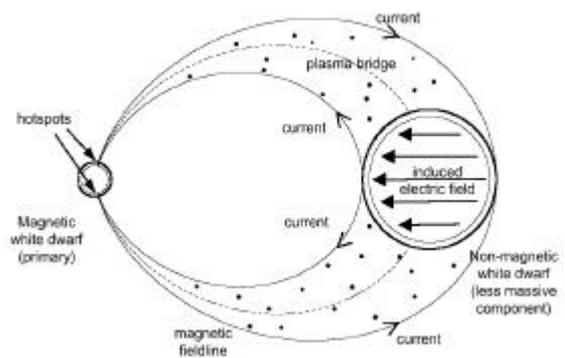

Fig. 4. A schematic illustration of a unipolar inductor consisting of a magnetic and non-magnetic white dwarf pair in a close binary orbit. [from Ref. 61].

In fact the generation of electric current between the magnetic object and non-magnetic body result in heating of the atmosphere/surface of the magnetic object where the current touches it and would result in production of emission in the polar region of the magnetic object. This could be one way of revealing a planet around a white dwarf that is difficult to detect otherwise [60].

Ref [61] proposed that binary stars consisting of a magnetic and a non-magnetic white dwarf can also be a cosmic unipolar inductor (cf. Fig. 4). In this model the luminosity is caused by resistive heating of the stellar atmospheres due to induced currents driven within the binary. This source of heating is found to be sufficient to account for the observed x-ray luminosity of the RX J1914+24, and provides an explanation for its puzzling characteristics [61].

Close binaries of this type can have short period and secondaries larger than the planet-size bodies. Provided that the spin of the magnetic white dwarf and the orbital rotation are not synchronized (so that there is a relative motion between the secondary and the magnetic field lines of the primary) and that the density of the plasma between the white dwarf is high enough, the unipolar inductor will operate.